\documentclass[12pt]{article}

\usepackage{amsmath,amssymb}
\usepackage{authblk}
\usepackage{graphicx}

\title{ Ising Game on Graphs}
\date{}
\author{Andrey Leonidov$^{(a,b)}$}
\author{Alexey Savvateev$^{(b,c,d)}$}
\author{Andrew G. Semenov$^{(a,e,f)}$}
\affil{{\small 
(a) P.N. Lebedev Physical Institute, Moscow, Russia\\
(b) Moscow Institute of Physics and Technology, Dolgoprudny, Russia\\
(c) Central Economics and Mathematics Institute, Moscow, Russia\\
(d) Adygea State University, Maykop, Russia\\
(e) National Research University Higher School of Economics, Moscow, Russia
(f) Skolkovo Institute of Science and Technology, Moscow, Russia
 }}

\begin{document}
\maketitle

\begin{abstract}
Static and dynamic equilibria in noisy binary choice (Ising) games on complete and random graphs in the annealed approximation are analysed. Two versions, an Ising game with interaction term defined in accordance with the Ising model in statistical physics and a reduced Ising game with a customary definition of interaction term in game theory on graphs, are considered.  A detailed analysis of hysteresis phenomenon shaping the pattern of static equilibria based on consideration of elasticity with respect to external influence is conducted. Fokker-Planck equations describing dynamic versions of the games under consideration are written and their asymptotic stationary solutions derived. It is shown that domains of parameters corresponding to the maxima of these probability distributions are identical with the corresponding hysteresis ranges for static equilibria. Same result holds for domains defining local stability of solutions of the evolution equations for the moments. In all the cases considered it is shown that the results for the reduced Ising game coincide with those obtained for the Ising game on complete graphs. It is shown that for s special case of logistic noise the results obtained for static equilibria for the Ising game reproduce those in the Ising model on graphs in statistical physics.


\end{abstract}

\newpage

\section{Introduction}

Taking into account direct not market-mediated economic/social interactions, i.e. a direct dependence of agent's utilities on the (expected) actions of fellow agents, is believed to be of crucial importance for taking into account an observed heterogeneity of economic/social multi-agent systems in describing their dynamical evolution and equilibria. The corresponding literature that contains many practical applications of this idea is covered in the review papers \cite{blume2001interactions,durlauf2010social}. Particularly interesting consequences of such interactions are emergent nontrivial aggregate properties of these systems such as phase transitions. 

At the same time the direct dependence of agent's utilities on the (expected) actions of other agents is, by default, a key ingredient of game theory. It is therefore natural to look at the consequences of economic/social interactions in terms of emergent  of characteristic regimes of evolutionary games and related aggregate properties  of game-theoretic equilibria.

One of the simplest settings in which effects of interactions, combined with those of noise and underlying topology, is a problem of noisy binary choice on graphs. There exists a significant literature on it with utility of a choice containing private and social components where the former depends on idiosyncratic fixed characteristics and noise and the latter depends on (expected) choices made by the agent's neighbours. The resulting description depends on the assumptions on the form of this interaction, topology of a graph and specification of noise. Static equilibria and corresponding myopic dynamics for noisy binary choice problem on complete graph were considered, for linear-quadratic utility and arbitrary noise distribution, in  \cite{brock2001discrete,blume2003equilibrium}.  
The particular case of linear-quadratic utility and logistic noise was considered for several topologies, including complete graph, star and circular ones\footnote{The case of the circular topology was also considered in the influential early paper \cite{glaeser1996crime}.} in \cite{ioannides2006topologies}. One of the main results of this literature is in describing nontrivial emerging aggregate properties of equilibria which in some special cases resemble or coincide with description of phases and phase transitions in statistical physics, see e.g. \cite{salinas2001introduction,Nishimori2011}. Of particular relevance to the present paper is an analysis of critical phenomena on graphs \cite{dorogovtsev2008critical}. An important analogy is here the one between the noisy binary choice problem in game theory and the Ising model (a particular model for magnets in which each node is characterized by one of the two possible orientations of spin) in statistical physics.  This is why it is natural to call a binary choice game an Ising game, a definition first given in \cite{galam2010ising}. An inspiring discussion of parallels between noisy discrete decision problems in multiagent systems and statistical physics can be found in \cite{bouchaud2013crises}, see also  \cite{durlauf2018statistical,durlauf1999can}. One particularly interesting phenomena analysed using methods of statistical physics are transitions between degenerate maxima of utility, see a discussion of a standard case in \cite{bouchaud2013crises} and an analysis for a self-excited Ising game in \cite{antonov2021self,antonov2023transition}. An interesting aspect of Ising games is a possibility of taking into consideration strategic forward-looking behaviour of agents leading to strategic cooling phenomenon \cite{leonidov2022strategic}.

As has been already mentioned, a natural language for developing a consistent description of the effects of economic/social interactions of a group of agents is provided by game theory in which interdependence of agent's strategies is built in by construction through their utilities. The goal is then to describe the dynamical evolution of such systems and the corresponding game-theoretic equilibria. The randomness present in agent's decisions makes it necessary to describe these evolution regimes and equilibria in probabilistic terms. Of particular relevance to the present study is a particular version of Bayes-Nash mixed strategies equilibria  - the Quantal Response Equilibria (QRE) \cite{mckelvey1995quantal,mckelvey1996statistical,goeree2016quantal}. An important result of \cite{mckelvey1995quantal} was in deriving explicit equations for equilibrium probability measure that were subsequently used for describing many experimental game-theoretic results \cite{goeree2016quantal}. 

Static equilibria in the Ising game on complete graphs with arbitrary noise were studied in \cite{brock2001discrete,blume2003equilibrium} and the corresponding dynamical mean field equations and an explicit construction of asymptotic equilibrium probability were described in \cite{blume2003equilibrium}. General equations defining static equilibria in the Ising game on graphs with arbitrary topology and general noise and identifying them as QRE ones and generic master equation for the system probability density were presented in   \cite{leonidov2019quantal,leonidov2020qre} \footnote{Let us note that although the relevance of QRE in the context of discrete choice models was already mentioned in the literature, see e.g. \cite{durlauf2010social,ioannides2006topologies}, the explicit interrelation was first quantified in \cite{leonidov2019quantal,leonidov2020qre} .}.  

In the present paper we expand the analysis in \cite{blume2003equilibrium,leonidov2019quantal,leonidov2020qre} in several directions. 

First, we consider two versions of noisy binary games differing by normalisation of the interaction term for graphs with arbitrary topology. In the first version termed the Ising game this normalisation is the same as in the Ising model on graphs in statistical physics \cite{dorogovtsev2008critical,dorogovtsev2002ising,leone2002ferromagnetic}. In the second one, termed the reduced Ising game, the interaction terms contains a multiplier equalling the inverse degree of the corresponding node. This normalisation is standard in game theory on graphs \cite{ioannides2006topologies,goyal2012connections} and does correspond to that in the nonlinear noisy voter's model \cite{peralta2018analytical}.

Second, we give a detailed analysis of hysteresis domain characterising static equilibria in the Ising and reduced Ising game for arbitrary noise on complete and random graphs based on the analysis of elasticity with respect to external influence - the quantity analogous to magnetic succeptibility in the Ising model.

Third, we analyse solutions of the Fokker-Planck equations describing dynamic versions of the Ising and reduced Ising games for arbitrary noise on complete and random graphs, identify conditions determining maxima of their stationary asymptotic solutions and analyse conditions for their stability. 
 
The plan of the paper is the following. 

In Section \ref{s:ImIg} we provide a condensed description of the Ising model in statistical physics relevant for the subsequent considerations in subsection \ref{ss:Im} and a general description of Ising and reduced Ising games in subsection \ref{ss:Ig}.

In Section \ref{s:static_game} we first provide a general description of static equilibria in the Ising and reduced Ising games on graphs. We proceed by giving a detailed analysis of static equilibria on complete graphs in subsection \ref{ss:secg}. We show that the results for the Ising and reduced Ising games on complete graphs are equivalent.  The results are compared with those obtained in the Ising model in statistical physics.    
 In subsection \ref{ss:serg} we analyse  static equilibria in the Ising and reduced Ising games on random graphs considered in the annealed approximation. We show that for the special choice of noise the results for the Ising game coincide with those obtained for the Ising model on graphs while those for the reduced Ising game are the same as in the Ising and reduced Ising games on complete graphs.

In Section \ref{s:evolutionary_game} we first provide a general description of dynamical Ising and reduced Ising games on graphs. We proceed by providing in subsection \ref{ss:decg} a detailed description of the Ising and reduced Ising games on graphs based on analysis of the Fokker-Planck equation derived from the master equation for the probability distribution of the average choice.  The dynamical evolution of the Ising and reduced Ising games on complete graphs is studied in subsection \ref{ss:decg} and on random graphs in the annealed approximation in subsection \ref{ss:derg}. We show that in all cases the stationary configurations coincide with the static equilibria described in Section \ref{s:static_game}. Conditions determining the maxima of equilibrium distributions are shown to coincide with those defining the hysteresis domain in the static games.

In Section \ref{s:conclusion} we conclude.

\section{Ising model and Ising games}\label{s:ImIg}

\subsection{Ising model}\label{ss:Im} Bridging, where possible, between results in game theory and statistical physics is of significant importance for understanding  large-scale properties of games of many agents. The particular noisy binary choice games considered in the paper are closely related to a generalized Ising model considered in statistical physics and machine learning, see e.g. \cite{Nishimori2011,nguyen2017inverse}. Let us therefore recall that a version of the Ising model on graphs in statistical physics related to the Ising game considered below is defined as a set of $N$ binary variables (spins) $\{ \sigma_i = \pm 1\}$ placed in vertices of a unoriented graph ${\cal G}$ influenced both by external fields and their neighbours so that the energy of a configuration $(\sigma_1, \dots, \sigma_N)$ reads
\begin{equation} \label{esf}
E (\sigma_1, \dots, \sigma_N) = - \left[  H \sum_{i=1}^N \sigma_i  + J \sum_{(ij)}  \sigma_i \sigma_j\right],
\end{equation}
where $H$ is an external magnetic field, $J>0$ is a coupling constant reflecting the interaction strengths and the summation in the second term is over all edges of the graph ${\cal G}$ so that
\begin{equation}
\sum_{(ij)}  \sigma_i \sigma_j = \sum_{i < j}  a_{ij} \sigma_i \sigma_j
\end{equation}
where $a$ is the adjacency matrix  of ${\cal G}$ with $a_{ij}=1$ corresponding to an edge $j \to i$ and $a_{ij}=0$ otherwise.
 
For a given spin $\sigma_i$ the energy of its interaction with the environment is
\begin{equation} \label{eisf}
E_i (\sigma_i) = - \left[  H  + J \sum_{ (ij) } \sigma_j \right] \sigma_i.
\end{equation}
In the presence of thermal noise the probability $p_{\sigma_i}$ of observing the state $\sigma_i$ depends only on $E_i (\sigma_i)$:
\begin{equation}
p_{\sigma_i} = \frac{e^{-\beta E_i (\sigma_i)}}{e^{-\beta E_i (\sigma_i)} + e^{-\beta E_i (-\sigma_i)}}
\end{equation}
where  $\beta=1/T$ is an inverse temperature. For the complete graph in the limit of large $N$
\begin{equation}\label{locprobIsing}
p(\sigma_i) = \frac{e^{(\beta H + \beta J m) \sigma_i}}{e^{(\beta H + \beta J m) \sigma_i}+e^{-(\beta H + \beta J m) \sigma_i}}, \;\; m = \frac{1}{N} \sum_i \sigma_i
\end{equation}
In turn, the probability $p(\sigma_1, \dots, \sigma_N)$ of a given configuration $(\sigma_1, \dots, \sigma_N)$ also depends only on its energy and equals
\begin{equation}\label{tnoise}
p(\sigma_1, \dots, \sigma_N) = \frac{e^{- \beta E (\sigma_1, \dots, \sigma_N)}}{Z}, \;\;\; Z=\sum_{\sigma_1, \dots, \sigma_N} e^{- \beta E (\sigma_1, \dots, \sigma_N)},
\end{equation}
The equilibrium configurations $(\sigma_1, \dots, \sigma_N)$ are those minimizing the free energy $F = - T \ln Z$. For the Ising model on the complete graph these are described by the Curie-Weiss equation  describing phase transitions in magnetics, see e.g. \cite{salinas2001introduction,Nishimori2011}
\begin{equation}\label{CWPh}
m = \tanh \left[ \beta H + \beta J m \right], \
\end{equation}

Static equilibria of the Ising model on uncorrelated networks with arbitrary degree distribution $\pi_k$ were considered in \cite{dorogovtsev2002ising,leone2002ferromagnetic}, see also a review \cite{dorogovtsev2008critical}.   In particular, it was shown that in the mean field limit the order-disorder phase transition point changes, as compared to the complete graph, as
\begin{equation}
\beta J = 1  \;\;\; \rightarrow \;\;\; \beta J = \frac{{\mathbb E}_\pi [k]}{{\mathbb E}_\pi [k^2]}
\end{equation}

A dynamical description of the Ising model is constructed in such a way that static equilibria are the asymptotic stationary configurations of a temporal evolution. There is no unique way of choosing a microscopic mechanism driving this evolution. Some popular choices are described in e.g.   \cite{salinas2001introduction,krapivsky2010kinetic}. The simplest mechanism considered in \cite{glauber1963time} suggests that dynamical evolution is driven by individual spin flips. For the Ising model on the compete graph the corresponding Glauber flip probability reads \cite{salinas2001introduction,krapivsky2010kinetic} 
\begin{equation}\label{glauber}
p (s \to -s) = \frac{1}{2} \left[ 1-s \tanh [\beta H + J m] \right].
\end{equation}

 Kinetics of the Ising model on random regular graphs in the annealed approximation was considered in \cite{jkedrzejewski2017kinetic}.

\subsection{Ising and reduced Ising games}\label{ss:Ig} Let us now define two versions of noisy binary choice games considered in the paper, an Ising game (IG) and reduced Ising game (RIG). The games are played by $N$ agents placed in the vertices of an unoriented graph ${\cal G}$. Each agent is equipped with two alternative pure strategies parametrised by $s_i = \pm 1$, $i=1, \cdots, N$. 

Static equilibria in noisy binary choice games result from agent's consideration of expected utilities characterising particular strategy choices of their neighbours.  Expressions for such expected utilities ${\mathbb E} \left[U_i (s_i)\right]$ of choosing a strategy $s_i$ for an agent $i$ for any {\it given} set of neighbour's strategies  $\{ s_j \}$, $\{ j \in {\cal V}_i \}$ where ${\cal V}_i$ is a set of indices of the first neighbours of a node $i$, for the two games under consideration read
\begin{eqnarray}\label{equation:utility_1}
{\mathbb E}_i \left[U^{\rm IG}_i (s_i) \right] &  = &  \left[ H  + J \sum_{j \in {\cal V}_i}  a_{ij}  {\mathbb E}_i \left[s_j \right] \right ] s_i  + \epsilon_{s_i} \label{uig}\\
{\mathbb E}_i \left[U^{\rm RIG}_i (s_i) \right]& = &  \left[ H +   \frac{J}{k_i } \sum_{j \in {\cal V}_i} a_{ij}   {\mathbb E}_i \left[ s_j \right] \right ] s_i  + \epsilon_{s_i} \label{urig}
\end{eqnarray} 
where ${\mathbb E}_i \left [ \dots \right ]$ is an expectation value as perceived by an agent $i$, $H$ is  a (common) bias towards choosing $s_i = {\rm sign} (H)$,   $J>0$ is a  constant reflecting the strength with which agents influence each other , $k_i = \vert  {\cal V}_i \vert$ is a degree of the node $i$. The variables $ \epsilon_{s_i} $ are strategy - dependent idiosyncratic random utility shifts visible only for an agent $i$ generated by some probability distribution $f(\epsilon_{s_i} \vert \beta)$ assumed to be the same for all  agents, where $\beta$ is a  scale characterising the noise level. It is assumed that for a graph under consideration $k_i = \vert  {\cal V}_i \vert \geq 1$ for all the nodes and that the probability distribution for $\{ \epsilon_{s_i} \}$ is a common knowledge. It is understood that when taking a decision both $\epsilon_{s_i}$ and $\epsilon_{-s_i}$ are known to an agent $i$.

The two utilities in \eqref{uig} and \eqref{urig} differ by rescaling the interaction contribution by $1/k_i$ in \eqref{urig}.  The expression  for $U^{\rm IG}_i (s_i)$ in \eqref{uig} has obvious parallels with that  for $E_i (\sigma_i)$  in \eqref{eisf} explaining the name chosen for this version of the game.  It is also important to mention that the RIG choice  \eqref{urig} is a common one  in  game-theoretic literature, see e.g. \cite{ioannides2006topologies} and references therein \footnote{In fact, as will be seen below, RIG corresponds to a game-theoretic version of a nonlinear noisy voter model, see e.g.  \cite{peralta2018analytical}.}. 

Let us stress that in contrast to equilibria in the generalized Ising model, Bayes-Nash game - theoretic equilbria in IG and RIG constructed below in Section \ref{s:static_game} result from constructing appropriate individually optimal decisions and not through maximisation of some common utility.  This difference is, in particular, reflected in the treatment of noise. The probabilistic measure corresponding to thermal noise in Eq.~\eqref{tnoise} is global, i.e. depends on the state $(\sigma_1, \dots, \sigma_N)$ of the whole system and the amount of noise is controlled by the single parameter $\beta=1/T$. On the contrary, probabilistic measures related to  random  utility shifts  $\{ \epsilon_{s_i} \}$ in (\ref{uig},\ref{urig}) are independent for different agents and, therefore, purely local. Moreover, the noise scale   $\{ \beta \}$ in the distribution functions $f(\epsilon_{s_i} \vert \beta)$ can also be different for different agents \cite{bouchaud2013crises,durlauf2018statistical}.  

Let us also mention that in statistical physics on graphs an adjacency matrix $\{ a_{ij} \}$ is necessarily symmetric (interaction energy is  link - related  ) while in games on graphs this is not required (utility is  node-related). However, in the present paper we will not consider binary choice games on directed graphs.

 In what follows we will use a condensed notation for local utilities in both games:
\begin{equation}\label{eugen}
{\mathbb E}_i \left[ U_i (s_i) \right]  =   \left[ H + \kappa_i   J \sum_{j \in {\cal V}_i}  a_{ij}  {\mathbb E}_i \left[ s_j  \right] \right ] s_i  + \epsilon^{(i)}_{s_i},
\end{equation}
where
\begin{itemize}
 \item for the Ising game $\kappa_i = 1/N$ for the game on complete graph and $\kappa_i = 1$ otherwise
 \item for the reduced Ising game $\kappa_i = 1/k_i$.
\end{itemize}

From Eq.~\eqref{eugen} one gets for the probability $p_{s_i}$ of choosing the strategy $s_i$ by an agent $i$:
\begin{equation}\label{psgen}
p_{s_i} = F \left[ 2 \beta \left( H + \kappa_i   J \sum_{j \in {\cal V}_i}  a_{ij}  {\mathbb E}_i \left[ s_j \right] \right) s_i \right]
\end{equation}
where
\begin{equation}\label{defF}
F (x \vert \beta_i) = {\rm Prob} \left[ \epsilon_{-s_i} - \epsilon_{s_i} < x \right] \equiv \int^x dx' \Phi (x' \vert \beta)
\end{equation}

The RIG version of the equation \eqref{psgen} coincides with the expression for the choice probability used in the nonlinear voter's game, see e.g.  \cite{peralta2018analytical}.

\section{Static equilibria}\label{s:static_game}

Expectation (epistemic, quantal response) equilibrium \cite{goeree2016quantal,mckelvey1995quantal,mckelvey1996statistical} is defined as a set of mixed strategies satisfying the consistency condition
\begin{equation}\label{eqcons}
{\mathbb E}_i \left[ s_i  \right]  = {\mathbb E}_j \left[ s_i  \right] = {\mathbb E}^{\rm eq} \left[ s_i  \right] ,  \;\; \forall i,j
\end{equation}
From Eqs.~(\ref{psgen},\ref{eqcons}) one gets \cite{leonidov2019quantal,leonidov2020qre} the following system of equilibrium - defining equations for $\{  {\mathbb E} \left[ s_i  \right]  \}$:
\begin{equation}\label{eqgen}
{\mathbb E}^{\rm eq} \left[ s_i \right] = 2 \left (2 \beta \left[ H + \kappa_i   J \sum_{j \in {\cal V}_i}  a_{ij}  {\mathbb E}^{\rm eq}  \left[ s_j  \right] \right ]   \right) -1
\end{equation}

In what follows  the games on random graphs are considered  in configuration model  \cite{newman2018networks} in the annealed approximation \cite{bianconi2007entropy} so that\footnote{This approximation was also used in the preceeding work \cite{leonidov2019quantal,leonidov2020qre}. } 
\begin{equation}
a_{ij} \;\; \to \;\; a_{ij}^{\rm ann} = \frac{k_i k_j}{N \langle k \rangle}
\end{equation}

Let us also define the following convenient log-odds parametrisation of the distribution function $F(x)$ introduced, e.g., in \cite{blume2003equilibrium}
\begin{equation}\label{logodds}
F(x) = \frac{e^{g(x)}}{1+e^{g(x)}} \;\; \leftrightarrow \;\;  g(x) = \log \frac{F(x)}{1-F(x)} 
\end{equation}
that will be used below.  

\subsection{Static equilibria: complete graph }\label{ss:secg}

Let us first consider the Ising game on a complete graph. In this case all nodes are equivalent so that
\begin{equation}
m^{\rm eq}_i = {\mathbb E}^{\rm eq} \left[ s_i \right] = m^{\rm eq} \;\; \forall i
\end{equation}
The equilibrium-defining equation \eqref{eqgen} takes the form
\begin{equation}\label{equation:QRE_av_cg}
m^{\rm eq} = 2F \left( 2 \beta H + 2 \beta J m^{\rm eq} \right) - 1 \equiv \tanh \left[ \frac{1}{2} g(2 \beta H + 2 \beta J m^{\rm eq})\right]
\end{equation}
i.e. the generalised Curie-Weiss equation obtained in  \cite{brock2001discrete,blume2003equilibrium} where in the second equality the parametrisation \eqref{logodds} was used. In the particular case of Gumbel noise (see Table 1 below) for which $g(x) =  x$ we have
\begin{equation}\label{eqprobIGcg}
p_{s_i} =  \frac{e^{(\beta H + \beta J m) s_i}}{e^{(\beta H + \beta J m) s_i}+e^{-(\beta H + \beta J m) s_i}}
\end{equation}
and
\begin{equation}\label{CWI}
m^{\rm eq}  =  \tanh \left[ \beta (H + Jm^{\rm eq} ) \right],
\end{equation}
which are equivalent to the local probability formula \eqref{locprobIsing} and the Curie-Weiss equation \eqref{CWPh} for the mean-field Ising model in statistical physics describing phase transitions in magnetics. This is in fact quite surprising, because Eq.~\eqref{equation:QRE_av_cg} describes competitive Bayes-Nash equilibrium realising specific agreement between maximisation of individual utilities  that is generally not related to any global optimisation procedure while the Curie-Weiss equation \eqref{CWPh} describes extrema of the free energy characterising the system as a whole.

The structure of solutions of \eqref{equation:QRE_av_cg} can be understood from considering the elasticity
\begin{equation}\label{succ_cg}
\chi (m^{\rm eq} \vert H) =\frac{\partial m^{\rm eq}}{\partial H} =  \frac{ \beta  \Phi (2 \beta H + 2 \beta J m^{\rm eq}) }{\left[1 - 4 \beta J  \Phi (2 \beta H + 2 \beta J m^{\rm eq}) \right]}. 
\end{equation}
From Eq.~\eqref{succ_cg}  we see that the structure of solutions of \eqref{equation:QRE_av_cg} is such that there is a transition between the high noise/weak interaction and low noise/strong interaction phases at
\begin{equation}
(\beta J)_{\rm crit} = C_d = \frac{1}{4 \Phi(0)}
\end{equation}
 The values of $C_d$ for three characteristic noise distributions (Gumbel, Gauss and t-Student) are given in Table 1:
\begin{center}
\begin{tabular}{ |c|c| } 
 \hline
 Noise distribution & $C_d = 4 \Phi(0)$ \\  \hline 
 $\phi_{\rm Gu} (x \vert \beta)= \beta e^{-\beta x + e^{- \beta x}}$ & $ 1 $\\  \hline
 $\phi_{\rm Ga} (x \vert \beta) = \frac{\beta}{\sqrt{2 \pi}} e^{- \frac{1}{2} (\beta x)^2}$ & $ \frac{\sqrt{\pi}}{2}$  \\  \hline
 $\phi_{\rm St} (x \vert \beta, \mu) = \frac{\beta}{\sqrt{\pi}} \frac{\Gamma \left( \frac{\mu+1}{2}\right)}{\Gamma \left( \frac{\mu}{2}\right)} \left[ 1 + (\beta x)^2)\right]^{-\frac{\mu+1}{2}}$ &  $ \frac{1}{4} \frac{\Gamma^2 \left( \frac{\mu}{2}\right)}{\Gamma^2 \left( \frac{\mu+1}{2}\right) } \frac{ (2 \mu )!!}{(2 \mu-1)!!}$  \\
 \hline
\end{tabular}
\end{center}
\begin{center}
\small{Table 1. Noise distributions and the correspondent constants $C_d$}
\end{center}

Let us note, that
\begin{itemize}
\item In the high-temperature phase $\beta J < C_d$ and, therefore, for all $(m^{\rm eq},H)$ 
\begin{equation}
1 - 4 \beta J  \Phi (2 \beta H + 2 \beta J m^{\rm eq}) > 0 \;\; \Rightarrow \;\; \chi (m^{\rm eq} \vert H) >0
\end{equation}
\item In the low - temperature phase $\beta J > C_d$ two regimes are possible:
\begin{eqnarray}
1 - 4 \beta J  \Phi (2 \beta H + 2 \beta J m^{\rm eq}) > 0 \;\; \Rightarrow \;\; \chi (m^{\rm eq} \vert H) >0 \label{chip} \\
1 - 4 \beta J  \Phi (2 \beta H + 2 \beta J m^{\rm eq}) < 0 \;\; \Rightarrow \;\; \chi (m^{\rm eq} \vert H) <0  \label{chim} 
\end{eqnarray}
The domains of these regimes are defined through finding the singular configurations $(m^{\rm eq}_{\rm sing},H_{\rm sing})$ such that
\begin{equation}
1 - 4 \beta J  \Phi (2 \beta H_{\rm sing} + 2 \beta J m^{\rm eq}_{\rm sing}) = 0 \;\; \Rightarrow \;\; \chi (m^{\rm eq} \vert H) = \infty
\end{equation}
\end{itemize}
Let us note that the regime in which $\chi (m^{\rm eq} \vert H) <0$ looks very unnatural. This feeling is supported by the fact that in the framework of dynamical game considered below it will turn out that such a regime is dynamically unstable.

\begin{figure}[ht]
	\begin{center}
		\includegraphics[width=0.7\linewidth, height=0.4\textheight]{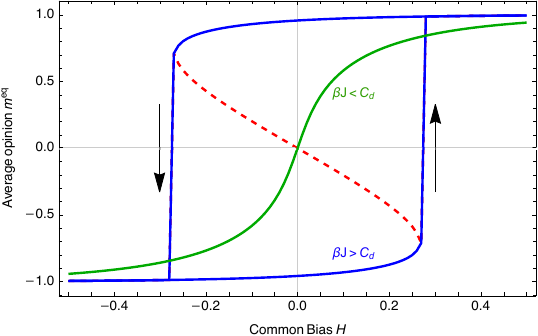}
	\end{center}
	\caption{The structure of solutions of \eqref{equation:QRE_av_cg}}
\label{figeq}
\end{figure}

The resulting structure of the solutions of \eqref{equation:QRE_av_cg} is illustrated in Fig.~\ref{figeq} and can be described as follows:
\begin{itemize}
\item At $\beta J< C_d$ (high noise/weak interaction phase) we have only one solution $m^{\rm eq} (H)$ with $\partial m^{\rm eq}/\partial H >0$ for all $H$  such that $m^{\rm eq} (H=0) = 0$, ${\rm sign} (m^{\rm eq} (H)) = {\rm sign} (H)$ and $\left. m^{\rm eq}(H) \right \vert_{H \to \pm \infty} \to \pm1$. The solution $m^{\rm eq} (H)$ does thus smoothly interpolate between the limits $m^{\rm eq}(H) \to -1$ for $H \to -\infty$ and  $m^{\rm eq}(H) \to 1$ for $H \to \infty$.
\item At $\beta J > C_d$ (low noise/strong interaction phase) there exist three solutions $\left( m^{\rm eq}_- (H),m^{\rm eq}_0 (H),m^{\rm eq}_+ (H) \right)$ parametrised by $m^*>0$ and $H^*>0$ such  that:
\begin{itemize}
\item The upper branch $m^{\rm eq}_+ (H)$ with $\partial m^{\rm eq}_+/\partial H >0$ exists for $H \geq - H^*$ so that  $m^{\rm eq}_+ (-H^*) = m^*$ and  $\left. m^{\rm eq}_+(H) \right \vert_{H \to \infty} \to 1$.
\item The lower branch $m^{\rm eq}_- (H)$ with $\partial m^{\rm eq}_-/\partial H >0$ exists for $H \leq H^*$ so that $m^{\rm eq}_- (H^*) = -m^*$ and  $\left. m^{\rm eq}_-(H) \right \vert_{H \to -\infty} \to -1$.
\item The intermediate branch $m^{\rm eq}_0 (H)$ with $\partial m^{\rm eq}_0/\partial H <0$ exists for $H \in [-H^*,H^*]$ so that it interpolates between $m^{\rm eq}_0 (-H^*) = m^*$ and $m^{\rm eq}_0 (H^*) = -m^*$.

The branches $m^{\rm eq}_\pm (H)$ realise the the regime in \eqref{chip} while the intermediate branch $m^{\rm eq}_0 (H)$ realises that in \eqref{chim}

\end{itemize} 
\end{itemize}

The parameters $m^*,H^*$ are determined by by looking for the values of $(m_{\rm sing},H_{\rm sing})$ such that $\partial m^{\rm eq}(H)/\partial H = \infty$  satisfying the following system of equations
\begin{eqnarray}\label{secsol1}
m_{\rm sing}^{\rm eq} & = & 2 F  [2 \beta H_{\rm sing} + 2 \beta J m_{\rm sing}^{\rm eq}] -1 \nonumber \\
1 & = &  4 \beta J \Phi [2 \beta H_{\rm sing} + 2 \beta J m_{\rm sing}^{\rm eq}] 
\end{eqnarray}
such that $(m_{\rm sing}^{\rm eq},H_{\rm sing}) = (m^*,-H^*)$ or $(m_s,H_s) = (-m^*,H^*)$, where $m^* \geq 0$ and $H^* \geq 0$.

For the Gumbel noise $g(x) =  x$ and the system \eqref{secsol1} can be solved analytically resulting in the following expressions:
\begin{eqnarray}
\frac{H^*}{J} & = &  \sqrt{\frac{\beta J - 1}{\beta J}} - \frac{1}{\beta J} {\rm atanh} \left( \sqrt{\frac{\beta J - 1}{\beta J}} \right) \nonumber \\
m^* & = & \sqrt{\frac{\beta J - 1}{\beta J}} 
\end{eqnarray}

For the reduced Ising game on a complete graph one has $\kappa_i = 1/(N-1)$ for all $i$ so that, in the considered limit of large $N$, its equilibria are defined by the same equation \eqref{equation:QRE_av_cg}.

\subsection{Static equilibria: random graphs}\label{ss:serg}

The only source of local heterogeneity taken into account in the configuration model of random graphs is related to a nontrivial distribution $\{ \pi_k \}$ over their degrees. A standard assumption (see e.g. \cite{goyal2012connections}) is then that vertices having the same degree have equivalent properties. In particular
\begin{equation}
m^{\rm eq}_{i (k)} = \left. {\mathbb E}^{\rm eq} \left[ s_i \right] \right \vert_{k_i = k} = m^{\rm eq}_k \;\; \forall i
\end{equation}
The equilibrium - defining equations \eqref{eqgen} then read\footnote{For the Ising game these equations were derived in  \cite{leonidov2019quantal,leonidov2020qre}.}:
\begin{equation}\label{eqgenrg}
m^{\rm eq}_k   =  2  F \left( 2 \beta H + 2 \beta J k \kappa_k \sum_{k'} \rho_{k'}  m^{\rm eq}_{k'} \right) -1 
\end{equation}
where the summation over $j$ is converted into the summation over $k$ and
\begin{equation}
\kappa_k = \frac{1}{k}, \;\;\; \rho_k = \frac{k  \pi_k}{{\mathbb E}_\pi [k]}. 
\end{equation}
Let us note that $\rho_k$  is a distribution of first neighbour's degree in the configuration model. It is convenient to rewrite the equilibrium - defining equations \eqref{eqgenrg}  introducing a weighted average
\begin{equation}
m_w = \sum_k \rho_k m_k.
\end{equation} 
We have 
\begin{eqnarray}\label{eqgenrg1}
m^{\rm eq}_k  & = &  2  F \left( 2 \beta H + 2 \beta J k \kappa_k m_w^{\rm eq} \right) -1 \label{eqgenrg1}  \\
m_w^{\rm eq} & = & 2 \sum_k \rho_k F \left( 2 \beta H + 2 \beta J k \kappa_k m_w^{\rm eq} \right) -1  \label{eqgenrg2}
\end{eqnarray}
From equations (\ref{eqgenrg1},\ref{eqgenrg2}) we see that the corresponding equilibrium is fully described by the properties of $m^{\rm eq}_w$ satisfying the closed equation \eqref{eqgenrg2}.

For the Ising game the structure of solutions of \eqref{eqgenrg2} can be understood from considering the elasticity
\begin{equation}
\chi_w (m^{\rm eq}_w \vert H)  =  \frac{\partial m_w^{\rm eq}}{\partial H} =  \frac{ 4\beta  \sum_k \rho_k \Phi (2 \beta H + 2 \beta J k m_w^{\rm eq}) }{\left[1 - 4 \beta J  \sum_k k \rho_k \Phi (2 \beta H + 2 \beta J k m_w^{\rm eq}) \right]}  \label{succ_rgmw}. 
\end{equation}

From Eq.~\eqref{succ_rgmw} we see that the structure of solutions of \eqref{eqgenrg2} is such that there is a transition between the high noise/weak interaction and 
low noise/strong interaction phases at
\begin{equation}\label{bjcrg}
(\beta J)_{\rm crit} = \frac{1}{4 \Phi(0)} \; \frac{{\mathbb E}_\pi[k]}{{\mathbb E}_\pi[k^2]} \equiv C_d  \frac{{\mathbb E}_\pi[k]}{{\mathbb E}_\pi[k^2]} 
\end{equation}
For the Gumbel noise $C_d=1$ and the expression \eqref{bjcrg} for the phase transition point coincides with the one obtained, within the same annealed approximation, in \cite{dorogovtsev2002ising,leone2002ferromagnetic}, see also \cite{dorogovtsev2008critical}.

For degree distributions with finite ${\mathbb E}_\pi[k^2]$ the structure of solutions of \eqref{eqgenrg1} is qualitatively the same as shown in Fig.~\ref{figeq}. If ${\mathbb E}_\pi[k^2]$ is infinite and, therefore, $(\beta J)_{\rm crit} =0$, the game will always stay in the low noise/strong interaction phase characterised by the presence of hysteresis..

The parameters $H^*_w$ and $m_w^*$ characterising the hysteresis domain in the low noise/strong interaction phase are determined by the system of equations for the singular configurations $(m^{\rm eq}_{w({\rm sing})},H_{\rm sing})$ 
\begin{eqnarray}
m_ {w({\rm sing})}^{\rm eq} & = & 2 \sum_k \rho_k F \left( 2 \beta H_{\rm sing} + 2 \beta J k m_{w({\rm sing})}^{\rm eq} \right) -1 \nonumber \\
1 & = & 4 \beta J  \sum_k k \rho_k \Phi (2 \beta H_{\rm sing} + 2 \beta J k m_{w({\rm sing})}^{\rm eq}) 
\end{eqnarray}
that has two solutions $(m_{w({\rm sing})},H_{\rm sing})=(m^*_w,-H^*_w)$ and $(m_{w(s)},H_s)=(-m^*_w,H^*_w)$ defining the borders of the hysteresis region.

For the reduced Ising game we get from \eqref{eqgenrg1}
\begin{equation}\label{eqgenrg3}
m^{\rm eq}_k   = m_w^{\rm eq}  = 2  F \left( 2 \beta H + 2 \beta J m_w^{\rm eq} \right) -1   
\end{equation}
i.e. $m^{\rm eq}_k = m^{\rm eq}_w \;\; \forall k$ and, in turn, the equation \eqref{eqgenrg3} for $m^{\rm eq}_w$ is the same as the equilibrium-defining equation \eqref{equation:QRE_av_cg} for the Ising game on complete graph.

\section{Dynamics}\label{s:evolutionary_game}

In the analysis of expectation equilibria in the static noisy binary choice games described in the previous section all solutions of the equilibrium-defining equations have the same status. A more nuanced analysis allowing to identify stable/unstable equilibria is possible through constructing dynamic noisy binary choice games having expectation equilibria as their stationary equilibrium configurations. The most natural approach is to consider dynamical games with many agents as population games in which  evolution is driven by birth-death type processes. For the Ising game on complete graphs such a population game was described in \cite{blume2003equilibrium,bouchaud2013crises}. In this section we give a detailed analysis of stability/instability of these equilibria in the Ising and reduced Ising games on complete and random graphs in the framework of the corresponding population games.

Analogously to the Glauber dynamics of the Ising model in statistical physics \cite{glauber1963time,salinas2001introduction,krapivsky2010kinetic} the version of a dynamical noisy binary choice game considered in \cite{blume2003equilibrium,bouchaud2013crises} and used in the present paper assumes that temporal evolution of the strategies space is driven by strategy reconsiderations of single agents so that at a given time moment $t$ in discrete time formulation or within an infinitesimally small temporal interval around $t$ in the continuous time formulation only one randomly chosen agent $i$ is given a right to consider a possible strategy change based on the corresponding {\it current} utilities $U_i (s_i \vert t)$. The expressions for these current utilities are {\it assumed} to read
\begin{equation}\label{dynut}
U_i (s_i \vert t )   =    \left[ H + \kappa_i  J \sum_{j \in {\cal V}_i}  a_{ij}  s_j (t) \right] s_i  + \epsilon_{s_i}
\end{equation}
where the set $\{ s_j (t) \}$is that of the {\it current} strategies of the neighbours and not the one of expectations of the agent $i$ with respect to his neighbour's choices as in the static case in Eq.~\eqref{eugen}. The expression \eqref{dynut} is thus a myopic version of the one in \eqref{eugen}. The corresponding probability of choosing the strategy $s_i$ is then given by an expression generalising Eq.~\eqref{psgen} to a dynamical setup\footnote{In order to avoid overloading of expressions below, here and in what follows we omit the explicit time index $t$ in the rhs of \eqref{eqpsdyn} and similar formulae.}:
\begin{equation}\label{eqpsdyn}
p_{s_i} = F \left[ 2 \beta \left( H  + \kappa_i  J \sum_{j \in {\cal V}_i}  a_{ij} s_j  \right) s_i \right]
\end{equation}

\subsection{Dynamical binary choice on complete graph}\label{ss:decg}

In this case all nodes are equivalent and a state $(s_1, \dots, s_N)$ of the system at time $t$ is fully described by the number $N_+ (t)$ of nodes equipped with the strategy $s=1$
The elementary events driving the evolution of the system are
\begin{equation}
N_+ \to N_+ \pm 1  \rightarrow   m \to m \pm \frac{2}{N}  
\end{equation}
where
\begin{equation}
m = \frac{1}{N} \sum_i s_i = \frac{N_+ - N_-}{N}
\end{equation}
 The probabilistic evolution of the system is fully described by the master equation on the probability distribution $P(m,t)$. Its evolution is driven by the rates
\begin{equation}
\omega \left( m \to m \pm \frac{2}{N} \right)   =   N_\mp \cdot p_\pm (m)  = \frac{N}{2} (1 \mp m) p_\pm (m)
\end{equation}
so that we get from Eq.~\eqref{eqpsdyn} the following expression for the probability $p_{s_i} (m)$ of choosing the strategy $s=1$:
\begin{equation}\label{probdyncg}
p_{s_i} (m) = F \left[ 2 \beta H + 2 \beta J m \right] = 1-p_{-s_i} (m)
\end{equation}
It is interesting to note that for the Gumbel noise the expression \eqref{probdyncg} coincides with the Glauber flip probability Eq.~\eqref{glauber}.

In what follows it will be convenient to introduce rates per node
\begin{equation}\label{redrates}
\omega_\pm (m)  =  \frac{1}{N} \; \omega \left( m \to m \pm \frac{2}{N} \right) = \frac{1}{2} (1 \mp m) p_\pm (m)  
\end{equation}

The probabilistic evolution of the system is fully described by the master equation on the probability distribution $P(m,t)$:
 \begin{eqnarray}\label{maseqcg}
  \frac{\partial P(m,t)}{\partial t}  & = & \omega \left( m - \frac{2}{N} \to m \right) P \left( m - \frac{2}{N},t \right)   + 
    \omega \left( m + \frac{2}{N} \to m \right)  P \left( m + \frac{2}{N},t \right) 
     \nonumber \\
   & - &   \left(\omega \left(m \to m - \frac{2}{N}  \right)  +  \omega \left(m \to m + \frac{2}{N}  \right)\right) P(m,t),
\end{eqnarray}
where, as previously discussed, we have assumed that within an infinitesimally small time interval it is possible to have only one strategy-changing event. 

In the considered limit of large $N$ the elementary step $2/N$ is small and, instead of solving the master equation  \eqref{maseqcg}, one can analyse the corresponding Fokker-Planck equation \cite{bouchaud2013crises,antonov2021self,antonov2023transition}
\begin{equation}\label{FPmcg}
 \frac{\partial P(m,t)}{\partial t}   =  - \frac{\partial}{\partial m} \left( 2 \left[ \omega_+(m) - \omega_-(m\right] P(m,t)\right) + 
 \frac{1}{N} \frac{\partial^2}{\partial m^2}  \left( 2 \left[ \omega_+(m) + \omega_-(m\right] P(m,t)\right)
\end{equation}

The stationary solution of the FP equation \eqref{FPmcg} is, with exponential accuracy
\begin{equation}
\frac{\partial P(m,t)}{\partial t} = 0 \;\; \rightarrow \;\; P^{\rm st}(m) \propto \exp \left[ N \int_0^m dm' 
\frac{ \omega_+(m') -\omega_- (m')}{ \omega_+(m') + \omega_- (m')} \right]
\end{equation}

 We have
\begin{equation}
P^{\rm st} (m)  \simeq e^{N u_{\rm FP} (m)},
\end{equation}
where
\begin{equation}\label{ufpw}
u_{\rm FP} (m)  =   \int_0^m dm' \; \frac{ \omega_+(m') -\omega_- (m')}{ \omega_+(m') + \omega_- (m')}  
\end{equation}
For general noise we have from (\ref{eqpsdyn},\ref{redrates})
\begin{eqnarray}
 u_{\rm FP} (m) & = &   \int_0^m dm' \; \frac{ \left( 2F[2 \beta H + 2\beta J m' ]-1\right) - m'}{1 + m' \left(2F[2 \beta H + 2\beta J m' ]-1 \right)} \nonumber \\
 & = & \int_0^m dm' \; \frac{\tanh \left(\frac{1}{2} g(2 \beta H + 2\beta J m' )\right) - m'}{1 - m'  \tanh \left(\frac{1}{2} g(2 \beta H + 2\beta J m' )\right)} 
\end{eqnarray}
The equilibrium state $m^s$ is naturally associated with that ensuring the maximal probability of occurrence: 
\[ 
m^s = {\rm argmax}_m [ P^{\rm st} (m)] =  {\rm argmax}_m [u_{\rm FP} (m)], 
\] 
The extrema $m^e$ of $u_{\rm FP} (m)$ satisfy the following equation
\begin{equation}\label{CWdcg}
u'_{\rm FP} (m^e)  =  0  \;\; \Rightarrow \;\;   m^e= 2F[2 \beta H + 2\beta J m^e ]-1
\end{equation}
which coincides with the equation \eqref{equation:QRE_av_cg} defining static equilibria and, therefore the structure of its solutions is the same as that described in the paragraph \ref{ss:secg} and can be reconstructed by analysing the properties of the elasticity
\begin{equation}\label{eldyncg}
\chi (m^e,H) = \frac{\beta \Phi (2 \beta H + 2\beta J m^e) }{1 - 4 \beta J \Phi (2 \beta H + 2\beta J m^e)}
\end{equation}
The solutions of \eqref{CWdcg} contain both maxima and minima. In contrast to the case of static equilibrium considered in the paragraph \ref{ss:secg} where all the solutions of \eqref{equation:QRE_av_cg} had the same status, here we can proceed to select only the maxima.

To select maxima of $u_{\rm FP} (m)$ one should find solutions $m^s$ of Eq.~\eqref{CWdcg} such that $u''_{\rm FP} (m^s)  <  0$. It is important to note that from the expression \eqref{ufpw} for $u_{\rm FP} (m)$ it is clear that, because of the obvious positivity of $\omega_+(m') + \omega_- (m')$, the sign of $u''_{\rm FP} (m^s)$ coincides with that of the $ [\partial_m (\omega_+(m) - \omega_- (m))]_{m_s} $, i.e. is fully determined by the first  term in the Fokker-Planck equation \eqref{FPmcg}. We have therefore
\begin{equation}\label{stabFPcg}
 u''_{\rm FP} (m^s)  <  0  \;\; \Rightarrow \;\; 4 \beta J \Phi (2 \beta H + 2\beta J m^s )  < 1 
\end{equation}
The condition for selecting the maxima is therefore the same as that ensuring $\chi (m^s,H)>0$. This holds for the solution branches $m^s_\pm (H)$ which, therefore, represent asymptotic stationary states. The intermediate branch $m^s_0 (H)$ for which $\chi (m^s_0,H)<0$ realises the minimum of $u_{\rm FP} (m)$ and thus does not correspond to an asymptotic  stationary state of the dynamics under consideration.

For the Gumbel noise
\begin{equation}
u_{\rm FP} (m)  =  \int_0^m dm' \; \frac{\tanh (\beta H +\beta J m') - m'}{1 - m' \tanh (\beta H + \beta J m') } 
\end{equation}

The asymptotic stationary states are defined by
\begin{eqnarray}
u'_{\rm FP} (m^s) & = & 0  \;\; \to \;\; \tanh (\beta H +\beta J m^s) - m^s= 0  \\
 u''_{\rm FP} (m^s) & < & 0 \nonumber  \;\; \to \;\;  (m^s)^2 > \frac{\beta J - 1}{\beta J} 
\end{eqnarray}
Let us note that in this particular case the range for $m_s$ does not depend on $H$.

Another way of selecting equilibrium configurations is to analyse their local stability by  considering the evolution of $\mathbb{E} [m]$. The corresponding equation can be derived from \eqref{FPmcg} or, equivalently, \eqref{maseqcg}.  It reads
\begin{equation}
\frac{d \mathbb{E} [m]}{dt} = - \mathbb{E} \left[ \omega_+(m) -\omega_- (m) \right] =  - \mathbb{E} \left[m - \tanh \left(\frac{1}{2} g(2 \beta H + 2 \beta J m) \right)\right]
\end{equation}

In the mean field approximation
\begin{equation}\label{mfdcg}
\frac{d \mathbb{E} [m]}{dt} = - \left( \mathbb{E} [m] - \tanh \left(\frac{1}{2} g(2 \beta H + 2 \beta J \mathbb{E} [m]) \right)\right).
\end{equation}
This equation can also be derived by constructing a population mean-field game, see \cite{blume2003equilibrium}. 
The  stationary points of \eqref{mfdcg} are described by
\begin{equation}\label{evavcg}
\frac{d \mathbb{E} [m]}{dt} = 0 \;\; \rightarrow \;\; \mathbb{E}^s [m] = \tanh \left(\frac{1}{2} g(2 \beta H + 2 \beta J \mathbb{E}^s [m]) \right)
\end{equation}
The stationary points $\mathbb{E}^s [m]$ are stable if
\begin{equation}
  \frac{\partial}{\partial  \mathbb{E} [m]} \left[ \mathbb{E} [m] - \tanh \left(\frac{1}{2} g(2 \beta H + 2 \beta J \mathbb{E} [m]) \right) \right]_{\mathbb{E}^s [m]}> 0
\end{equation}
which has the same form as \eqref{stabFPcg} and, therefore, the (locally) stable solutions of .\eqref{evavcg} are $\mathbb{E}^s_\pm [m]$.

The results for the reduced Ising game on complete graphs coincide with the above-described ones.

\subsection{Dynamical binary choice on random graphs}\label{ss:derg}

For the Ising game on random graphs all nodes with the same degree are equivalent and the state of the system at time $t$ is fully described by the set $\left(N^+_1, N^+_2, \dots,N^+_k, \dots \right)$, where $N^+_k$ is a number of nodes with degree $k$ equipped with the strategy $s=1$.
The elementary events driving the evolution of the system are
\begin{equation}
N^+_k \to N_+^k \pm 1  \rightarrow   m_k \to m_k \pm \frac{2}{N_k} \nonumber \\
\end{equation}
where
\begin{equation}
m_k = \frac{N^+_k - N^-_k}{N_k}
\end{equation}

The evolution of $P(m_k,t)$ and $P(m_w,t)$ is driven by the rates
\begin{equation}
\omega^{(k)} \left( m_k \to m_k \pm \frac{2}{N_k} \right)   =   N_k^\mp \cdot p_\pm^{(k)} (m_w)  = \frac{N_k}{2} (1 \mp m_k) p_\pm^{(k)} (m_w)
\end{equation}

In what follows it will be convenient to introduce rates per node
\begin{equation}
\omega_\pm^{(k)} (m_k,m_w)  =  \frac{1}{N_k} \; \omega^{(k)} \left( m_k \to m_k \pm \frac{2}{N_k} \right) = \frac{1}{2} (1 \mp m_k) p_\pm^{(k)} (m_w)  \end{equation}

The corresponding Fokker-Planck equations read
{\small
\begin{eqnarray}\label{FPmk}
 \frac{\partial P(m_k,t)}{\partial t}  & = & - \frac{\partial}{\partial m_k} \left( 2 \left[ m_k - \tanh \left(\frac{1}{2} g(2 \beta H + 2 \beta J k m_w) \right) \right] P(m_k,t)\right) \\
 &+&  \frac{1}{N_k} \frac{\partial^2}{\partial m_k^2}  \left( 2 \left[ 1 - m_k  \tanh \left(\frac{1}{2} g(2 \beta H + 2 \beta J k m_w) \right) \right] P(m_k,t) \right) \nonumber
\end{eqnarray}

\begin{eqnarray}\label{FPmw}
\frac{\partial P(m_w ,t)}{\partial t} & = &  - 
\frac{\partial}{\partial m_w} \left( \left[ m_w - \sum_k \rho_k \tanh \left( \frac{1}{2} g(2 \beta J k m_w) \right) \right] P(m_w ,t) \right) \\
&+&  \frac{1}{N} \frac{\partial^2}{\partial m_w^2} 
\left( \left[ \sum_k  k \rho_k   \left(1 - m_k  \tanh \left( \frac{1}{2} g(2 \beta J k m_w) \right) \right) \right] P(m_w ,t) \right)  
 \nonumber
\end{eqnarray}
}

Equation \eqref{FPmw} is a closed equation on $m_w$. Its stationary solution reads
\begin{equation}
P^{\rm st} (m_k )  \simeq e^{N u_{\rm FP} (m_w)}
\end{equation}
where
\begin{equation}\label{uFPmw}
u_{\rm FP} (m_w) = \int_0^{m_w} dm' \; 
\frac{ \sum_k \rho_k \tanh \left[ \frac{1}{2} g(2 \beta J k m') \right] - m'}{\sum_k  \pi_k \frac{k^2}{\langle k \rangle^2}   \left(1 - m_k  \tanh \left[ \frac{1}{2} g(2 \beta J k m') \right] \right)  }  
\end{equation}

Locations of the extrema $m_w^e$ of $u_{\rm FP} (m_w)$ depend only on the numerator in the right-hand side of Eq.~\eqref{uFPmw} and satisfy  the equation $u'_{\rm FP} (m_w^e)  =  0$. One has
\begin{equation}\label{extFPrg}
u'_{\rm FP} (m_w^e)  =  0  \;\; \to \;\;  m_w^e =  \sum \rho_k \tanh \left(\frac{1}{2} g(2 \beta H + 2\beta J k m_w^e )\right)
 \end{equation}
which coincides with the equation \eqref{eqgenrg2} defining static equilibria. The corresponding maxima are defined by $m_w^s = {\rm argmax}_{m_w} [u_{\rm FP} (m_w)] $ and are thus defined by
\begin{eqnarray}\label{stabFPrg}
 u''_{\rm FP} (m_w^s)  <  0   \;\; \to \;\;  \sum  k \rho_k \frac{ \beta J g' (2 \beta H + 2\beta J k m_w^s )}{\cosh^2 \left(\frac{1}{2} g(2 \beta H + 2\beta J k m_w^s )\right)}  > 1
\end{eqnarray}
From equations (\ref{extFPrg},\ref{stabFPrg}) we see that, in complete analogy with the above-considered case of complete graph, the condition \eqref{stabFPrg} identifying the asymptotic stationary states is equivalent to the condition of positive elasticity $\chi^s_w(m^s_w,H)$ which holds for the solution branches $m^s_{w \pm} (H)$.

A direct analysis of the stability of the stationary values of $m_w$ can be performed by considering the evolution of $\mathbb{E} [m_w]$ From \eqref{FPmw} it follows that
\begin{equation}
\frac{d \mathbb{E} [m_w]}{dt} =  - \mathbb{E} \left[m_w - \sum_k \rho_k \tanh \left(\frac{1}{2} g(2 \beta H + 2 \beta J k m_w) \right)\right] 
\end{equation}
In the mean field approximation
\begin{equation}
\frac{d \mathbb{E} [m_w]}{dt}  =  - \left( \mathbb{E} [m_w] - \sum_k \rho_k \tanh \left(\frac{1}{2} g(2 \beta H + 2 \beta J k \mathbb{E} [m_w]) \right)\right) 
\end{equation}
and, therefore, for the stationary solution one has
\begin{equation}\label{ssmw}
\frac{d \mathbb{E} [m_w]}{dt}  =  0 \;\; \rightarrow \;\; \mathbb{E} [m_w] = \sum_k \rho_k  \tanh \left(\frac{1}{2} g(2 \beta H + 2 \beta J k \mathbb{E} [m_w]) \right)
\end{equation}
The stationary solution \eqref{ssmw} is stable if
\begin{equation}
\frac{d}{d \mathbb{E} [m_w]} \left[ \sum_k \rho_k \tanh \left(\frac{1}{2} g(2 \beta H + 2 \beta J k \mathbb{E} [m_w]) \right) - {\mathbb E} [m_w] \right] >0
\end{equation}
which is equivalent to the condition $\chi_w (m^{\rm eq}_w \vert H)>0$ so that the stable solutions are ${\mathbb E}[m_w]_\pm (H)$.  

\section{Conclusions}\label{s:conclusion}

Let us summarise the main results obtained in the present study:

\begin{itemize}
\item Static and dynamic equilibria in the Ising and reduced Ising game, the latter equivalent to the game-theoretic version of the noisy nonlinear voter's model, we considered for arbitrary noise on complete and random graphs in the annealed approximation. 
\item In all the cases considered the results for the reduced Ising game were shown to be equivalent to the Ising game on complete graph.
\item A detailed analysis of hysteresis phenomenon characterising static equilibria in the Ising and reduced Ising games based on the analysis of properties of elasticity with respect to external influence was presented. In the particular case of logistic noise the results for the static equilibria were shown to be equivalent to those for the Ising model in statistical physics.
\item Based on the derived Fokker-Planck equations describing dynamical versions of the Ising and reduced Ising games maxima of asymptotic stationary probability distributions and stability of their solutions were analysed. It was shown that domains determining maxima of stationary distributions, domains of dynamical stability and hysteresis ranges turned out to be the same in all the cases considered.

\end{itemize}



\end{document}